\documentclass[CRMECA,Unicode,manuscript,a4paper]{cedram}

\usepackage[all]{xypic}
\usepackage[range-phrase = --,retain-unity-mantissa = false,exponent-product = \cdot]{siunitx}

\usepackage{hyperref}

\title{The ping-pong ball water cannon}

\author[B. Andreotti]{\firstname{Bruno} \lastname{Andreotti}}
\address{Laboratoire de Physique de l'ENS, UMR 8550 Ecole Normale
  Sup\'erieure -- CNRS -- Universit{\'e}~de~Paris --
  Sorbonne~Universit{\'e}, 24 rue Lhomond, 75005 Paris}
\email{bruno.andreotti@phys.ens.fr} \thanks{The authors thank the
  whole team of Physique Expérimentale, and Michael Berhanu. The first
  author is supported by ANR through grant Smart. The Physics
  department of Université de Paris also contributed by supporting its
  team's preparation for the International Physicists' Tournament.}
\author[W. Toutain]{\firstname{Wladimir} \lastname{Toutain}}
\address{Mati{\`e}re et Syst{\`e}mes Complexes, UMR 7057
  Universit{\'e} de Paris -- CNRS, 10 rue Alice Domon et L{\'e}onie
  Duquet, 75013 Paris, France.} \author[C.
No{\^u}s]{\firstname{Camille} \lastname{No{\^u}s}}
\address{Laboratoire Cogitamus, \url{https://www.cogitamus.fr/}}
\author[S. El Rhandour-Essmaili]{\firstname{Sofia} \lastname{El
    Rhandour-Essmaili}} \address[2]{foo} \author[G.
P{\'e}rignon-Hubert]{\firstname{Guillaume}
  \lastname{P{\'e}rignon-Hubert}} \address[2]{bar} \author[A.
Daerr]{\firstname{Adrian} \lastname{Daerr}} \address[2]{baz}
\email{adrian.daerr@univ-paris-diderot.fr}

\CDRGrant[ANR]{Smart}

\keywords{Scientific method,teaching of research, fluid mechanics, hydraulic catapult, shock wave, pressure focussing, liquid projection}

\subjclass{00X99}

\begin{abstract} 
  The course ``Phy Ex'' was created by Yves Couder in the Paris VII
  university to teach experimental physics through projects. In this
  article, we present this teaching method through a particular
  project that took place in the autumn semester 2019: the ping-pong
  ball water cannon. In this experiment, a glass containing water and a
  floating table tennis ball is dropped from some height to the
  ground. Following the impact, the ball is ejected vertically upwards
  at speeds that can be several times the impact speed. We report the
  student team's initial dimensional and order-of-magnitude analysis,
  and describe the successive experimental set-ups that showed (1)
  that free flight is essential for the phenomenon to occur, (2) that
  the order of magnitude of the ball ejection momentum is correctly
  predicted by a momentum balance based on integrating the pressure
  impulse during impact and (3) that making the ball surface more
  wettable, or stirring the liquid, drastically increases the momentum
  transfer. The proposed explanation, confirmed by direct high-speed
  video observations, is that the immersion depth of the ball
  increases during free fall due to capillary forces or vortex
  depression --- in the absence of buoyancy --- and that the enormous
  excess pressure on the bottom of the ball during impact drives the
  ball up towards its buoyancy equilibrium. The transfered momentum is
  sufficient to expel the ball at high velocity, very similar to the
  formation of liquid jets in collapsing cavities in liquids.
\end{abstract}

\begin{document}

\maketitle

\selectlanguage{french}
\section*{Version française abrégée}

Yves Couder a créé un enseignement de physique expérimentale par
projets, «Phy Ex», à l'université de Paris VII. Cet article présente
la méthode d'enseignement à travers le projet d'un binôme d'étudiants
au premier semestre 2019/2020: le canon à eau propulsant une balle de
ping-pong. Le projet consistait à identifier les mécanismes physiques
en jeu dans l'expérience suivante: on fait tomber verticalement un
verre en plastique contenant de l'eau avec une balle de ping-pong à sa
surface. Lorsque le verre percute le sol, la balle de ping-pong est
éjectée vers le haut à une vitesse pouvant largement dépasser la
vitesse de chute. Nous résumons l'analyse dimensionnelle et d'ordre de
grandeur faite par le binôme d'étudiants en amont des expériences.
Puis nous décrivons les montages successifs qui montrent (1) que la
phase de chute libre précédant l'impact est essentielle dans ce
phénomène, (2) que l'ordre de grandeur de la vitesse d'éjection est
correctement prédit par un bilan de quantité de mouvement basé sur
l'intégration du pic de pression résultant de l'impact, et (3) que
rendre la balle plus mouillante, ou que mettre l'eau en rotation avant
la chute, résulte en une augmentation importante de la quantité de
mouvement transmise à la balle. L'image physique ayant émergé au cours
de ce projet, et qui s'appuie également sur des images vidéo prises à
l'aide d'une caméra rapide, est que la profondeur d'immersion de la
balle augmente pendant la brève période de chute libre par l'action
des forces capillaires ou par la dépression au c{\oe}ur d'un vortex en
présence de rotation --- la poussée d'Archimède habituellement dominante
étant nulle pendant la chute --- et que la pression considérable sur la
partie immergée de la balle pendant la décélération au sol repousse la
balle vers son équilibre de flottaison. La quantité de mouvement
transmise dans ce processus est alors tellement importante que la
balle est expulsée vers le haut à grande vitesse, de la même manière
que l'effondrement d'une cavité à la surface d'un liquide conduit à la
formation d'un jet rapide au centre.

\selectlanguage{english}

\section{On the ``Phy Ex'' course at Université Paris VII}

\subsection{On the importance of teaching experimental projects in
  Physics}

Although people, in most industrialised societies, still trust science
far more than they (statistically) trust governments or enterprises,
in some areas like climate change, evolution or vaccination, the level
of public suspicion is high. The most obvious threat to a public
debate based on facts and scientific arguments, the ``post-truth'',
results from the ``destruction (\dots) of the effective, public social
norms and behaviours that the common search for truth
presupposes''\cite{castoriadis1979nouvelObs}. However, beyond the
gradual disappearance of a public space of thought and confrontation
based on shared objective standards for truth, the scientific world
faces another danger. Great confusion in the public debate results
from the belief that the Truth will naturally emerge from the free
competition of opinions and ideas, in an open marketplace of ideas —
called the \textsl{Catallaxy} by Hayek\cite{hayek2012law}. It is
taken for granted that the collective information processing in a
society is spontaneously efficacious in producing knowledge --- one must
assume more than any other human collective activity --- regardless of
the participants' training, education, expertise or even ability to
reason. Some of the advocates of Open Science and ``Science 2.0'',
amongst which think tanks and fake institutes like the Ronin
Institute, Center for Open Science, openscienceASAP, UK Open Data
Institute, PCORI or the Laura and John Arnold Foundation, to cite just
a few, promote the substitution of internet platforms, inspired from
social media, for old style scientific procedures, based on the
Humboldtian model. The discredit of science used as a political
strategy (by some actors in the fossil fuel industry, for instance, in
the face of climate change) should alarm scientists on the more subtle
calls for deregulation of scientific methods, claiming that
spontaneous order will magically solve the nuisances of contemporary
science, including the occasional corruption of private scientific
journals and the crisis of replicability.

It has always been part of scientific education to disentangle belief
and knowledge, to identify the proper contours of the scientific and
political spheres and to discuss the grey zone in between: expertise
and techniques. The threats to scientific thought now require
academics to pay particular attention to the place given in university
training to the scientific method and its specific rules of probation
and truthfulness. Experimenting with original teaching methods to do
this is all the more urgent since, taking advantage of the prevailing
credulity, more and more pressure groups are trying to claim the
`scientific' nature of their opinions, developed outside any
scientific method. The scientific method is based on the establishment
and publication of objectifiable facts (in physics, experimental
measurements and mathematical theories) published alongside a
discourse of proof and an extensive review of the scientific
literature on the subject.

The scientific method needs to be taught as early as possible, yet
much of it can only be learned through guided practice. How can
undergraduate students learn the scientific method when they do not
have a sufficient theoretical background to tackle open problems on
the forefront of science? Experience shows that pretending to
``investigate'' a problem whose solution is perfectly known is
tantamount to making science a dead language. Actually, most of the
french elite is formed in dedicated professional schools, known since
Napoleonic times as ``Grandes Écoles'', where students never encounter
a problem which does not present a clean, well known, analytical
solution. In the third year of the bachelor's degree at the physics
department of Paris VII university, the attempt at a more fruitful
approach consists in letting students conduct experimental research in
teams of two over the course of one semester, on a subject chosen from
a range of largely open-ended scientific problems in macroscopic
physics.

This presupposes suitable problems, and means. Students are thus led
to work on a scientific question by themselves, to design and
implement an experiment, to develop a model, to read primary
scientific sources, to practice peer review (\textsl{disputatio}), to
write down a draft including scientific evidence and reasoning. This
guided approach to research has been pioneered by Yves Couder: the
next section tells of the genesis, the initial rationale and the
evolution of this course over the past decades. In the second part of
this article, we will report on one of the experiments performed by a
group of students in the academic year 2019--2020.

\subsection{A short history of experimental physics projects (\textsl{Phy Ex})
  at Univ. Paris VII}

Yves Couder has created this experimental physics course (here after
referred to as \textsl{Phy Ex}) in an entirely different context, at
the creation of the Université Paris VII, born from the division of
Université de Paris. Université Paris VII, which had also gone by the
name of Université Paris Diderot in the past years, has recently
merged with Université Paris 5 to become Université de Paris, closing
a historic cycle. \textsl{Phy Ex} was inspired by the novel
pedagogical approach developed in the \textsl{Centre universitaire
  expérimental de Vincennes} after 1968. Yves Couder had met Michel
Juffé during the 1968 movement at the \textsl{Commission Nationale
  Interdisciplinaire} (CNID, rue d'Assas, Paris). From May to June
'68, under the presidency of Marc Hatzfeld, student in political
science, this group --- which included personalities such as Michel
Alliot, Pierre Bourdieu, Henri Cartan, Jacques de Chalendar, Hélène
Cixous, Jacques Monod and Laurent Schwartz --- produced a programmatic
document to change University, entitled ``Proposition pour de
nouvelles structures universitaires''. In 1972, at the
\textsl{Département de Sciences de l'éducation}, in Vincennes, an
innovative method (BETIS) was experimented by Guy Berger, Ruth Kohn,
Yves Couder, Michel Juffé and Antoine Savoie: between 30 and 40
undergraduate students, in groups of 5 or 6, where asked to produce an
interdisciplinary work on educational methods, based on book reading,
discussion and research. The BETIS research project would be awarded
from 0 to 10 credits --- out of the 30 required to obtain the
\textsl{Licence} (Batchelor) grade in three years --- depending on the
amount of work. This method favouring autonomy turned out to be so
time demanding that the students could hardly follow other courses. It
was one full university year out of three devoted to research and
\textsl{disputatio}. The role of the five teachers (a spinozist
philosopher, a physicist, a critical sociologist from the
institutional analysis school, an epistemologist practising the
participant observation, and a pedagogue) was to guide, to discuss, to
improve the quality of reasoning. Forming groups with students of all
three \textsl{Licence} years, having therefore different degree of
maturity in their disciplines, proved most effective. The project
oriented teaching method became official in 1976 and remained active
for decades. When Michel Juffé left Vincennes in autumn 1977, Yves
Couder stopped teaching there and started \textsl{Phy Ex} at Paris VII
university. The paper archives of all the experimental projects since
the creation of the course are conserved to date in the experimental
room.

The original motivation for \textsl{Phy Ex} was directly related to
the emancipation movement of the 1970's. The idea of a co-production
of knowledge by students immediately after attending University was a
reaction to both the conservative old style of Sorbonne bigwigs and to
the post-war development of a standardised mass university. It aimed
at breaking or at least re-examining the existing hierarchy between
teachers and students. Simultaneously, an experimental physics course
would promote practical knowledge against theoretical knowledge and
thus in principle add value to the popular do-it-yourself culture
(DIY, \textsl{bricolage}), against a cultural capital backdrop of
mathematics accused of reproducing inequalities by social
inheritance. Yves Couder himself, in his research, has defended the
possibility of pure experiments not initially motivated by any theory,
starting from minute table-top experiments. He had two simple
principles to let surprises emerge from an experimental set-up: turn
the button to the maximum, and ask yourself and others ``what happens
if''. For instance, Yves Couder's famous bouncing drop
experiment\cite{couder2005walking,couder2005bouncing} started in \textsl{Phy Ex} as a
Faraday instability experiment, using a sabre saw as shaking device.
How could he resist encouraging the students to push the sabre saw
vibration to the maximum, rather than to obediently reproduce curves
provided in the scientific article they were starting from.

Since the late seventies, social changes have profoundly modified the
perception of \textsl{Phy Ex}. Experimental projects in physics are
now taught in many universities. Projects are no longer associated
with innovative emancipatory teaching methods, but with mainstream
management control techniques. The current goal of \textsl{Phy Ex} has
conceptually changed a lot, although the teaching itself may be close
to its original form. The students, for the first time, are confronted
with the scientific method, and with knowledge less well established
and polished than what they are taught in regular courses. They
discover the existence of academic journals, of formal peer review of
papers. They see the social practices around science: for instance, in
\textsl{Phy Ex}, preliminary results are frequently discussed; it is
only progressively that the scientific matter is settled and that
knowledge coming from an experiment becomes trustworthy. They have to
identify key control parameters, to perform measurements, to
understand how an experiment is designed and built to solve a question.
As soon as they have experimental data, they are confronted to ethics
and intellectual virtues — starting with: am I allowed to remove a bad
data point? A permanent expectation since the pioneering times of
\textsl{Phy Ex} is to create the conditions of a particular
involvement of students in their experiments, an appropriation of
knowledge. The success of a project is judged primarily by the hedonic
reward provided by a resistant experiment that suddenly starts working
after some effort. As manual skill is decisive for this, \textsl{Phy
  Ex} reveals to dozens of students each year their ability to devise
and develop practical tools to understand a physical phenomenon. They
learn that nature can be made to reveal some of its inner workings
through some ingenious set-up, rather than through theoretical
calculus, and that an experiment is but a dialog where one has to
learn to ask the right questions and make sense of the answers. Hence
the pleasure students ordinarily report experiencing in the last few
weeks of the semester.

\subsection{The contemporary organisation of  \textsl{Phy Ex}}
Experimental projects in physics at Paris 7 university take 8 hours a
week (2 times 4 hours) over the course of a semester (12 weeks). Each
group is composed of 30 to 36 students, who work in pairs on 15 to 18
projects. A well equiped machine shop with one permanent lathe-mill
operator is assigned to \textsl{Phy Ex} to help setting up the
experiments. A second technician is responsible for the material
(especially fragile or expensive parts) and for the supervision of
computers. Three or four experienced academics are needed to supervise
the projects. Initially, around 25 projects are proposed by the
teachers, based on open problems or on a physics paper. An essential
rule is to check that there is at least one quantity to measure and
one control parameter, in order to ensure that an experimental curve,
at least, can be drawn. Open questions that are too qualitative, or
that involve fields rather than scalar quantities are rejected. There
is a collective check of prior feasibility. As subjects in acoustics,
hydrodynamics, capillarity, optics, electrostatics, elasticity,
granular material, \textsl{etc}, are easier to design, attention is
paid to avoiding many subjects in the same field, and to finding
subjects involving quantum mechanics, solid state physics,
astrophysics, and so on.

During the first hour, each subject in turn is described in two
minutes and one video-projected slide in front of all students. They
are then invited to sign up for up to three subjects, with a priority
on one of them. During one hour, we help resolving cases where several
groups would like to work on the same problem. It is important to
avoid initial frustration as much as possible. Then, the students
devote the first 5 time slots (20 hours) to the theoretical analysis
of their subject. They are asked to formalise a question in scientific
terms, sometimes starting from a problem posed in every day terms.
Most often, they have to read (for the first time) a scientific paper,
see the typical structure and the bibliography. Most of the time is
then devoted to reading chapters of books to learn elementary physics
that they need for the subject. In the specific case presented below,
the students had no background in either hydrodynamics or capillarity:
they only knew rigid body mechanics and hydrostatics. Very often, the
students are encouraged to perform the dimensional analysis of their
problem, which is usually far too complicated to be solved
theoretically. They need to think about the orders of magnitude
involved in the experiment and the way they will measure central
quantities. At the sixth sequence, all the groups present their
theoretical work and provide a short written report. They then work on
their experiment for the next 18 periods of 4 hours. At the end, they
are expected to obtain reliable experimental curves and to be able to
discuss them. In some cases, the comparison with the theory can be
pushed further. The students finally write a second report and prepare
an oral presentation of the whole project in front of all other
students and teachers.

In the rest of the article, we will report the project performed by
Sofia El Rhandour-Essmaili and Guillaume P{\'e}rignon-Hubert, with the
technical help of Wladimir Toutain, in autumn 2020.
 \begin{figure}[tbp]
\includegraphics{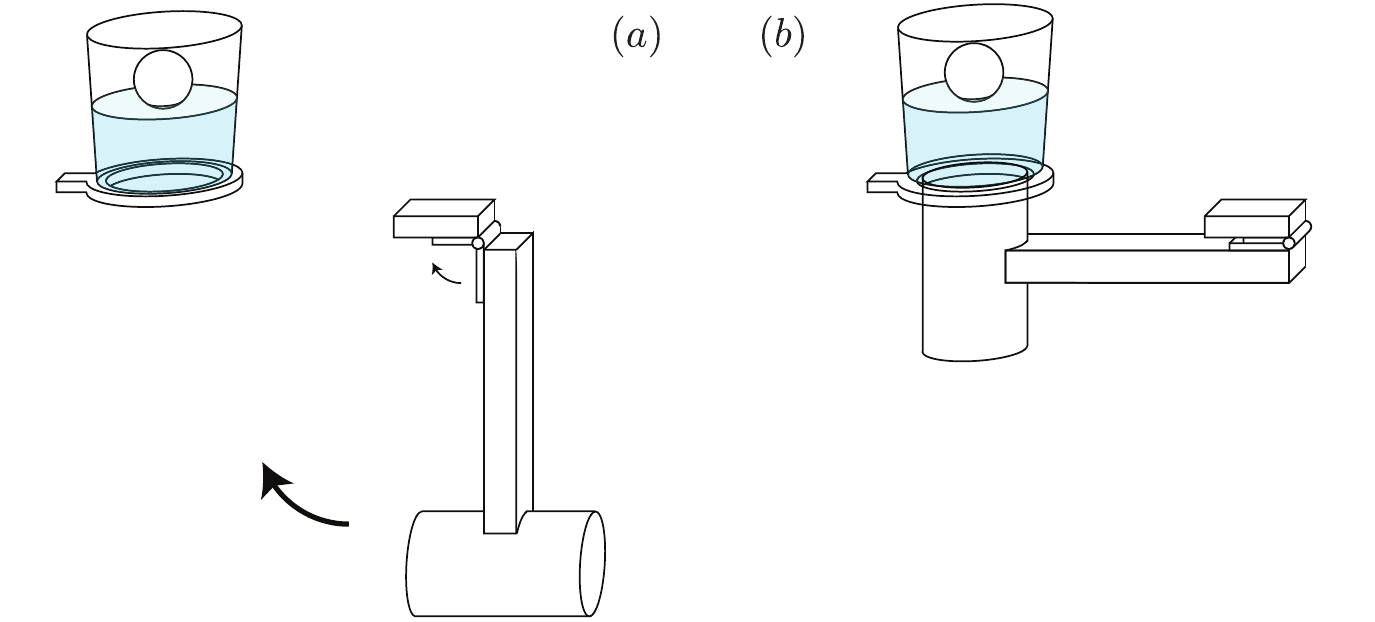}
 \caption{Schematic of the first experimental set-up, using a pendulum
   as a hammer to hit the bottom of the glass.}
 \label{fig:schematicHammer}
 \end{figure}

\section{Problem and theoretical developments}
\subsection{The water canon problem}
The project discussed here as an example was proposed as problem
number 1 in the International Physicists' Tournament (IPT) 2020 under
the title \textsl{Cumulative cannon}. The subject description pointed
to a Youtube video~\cite{youtube} where `Mr. Hacker' puts a table
tennis ball in a transparent plastic glass partly filled with water,
that he has previously set into rotation, and lets the glass fall to
the floor. Just after the impact with the ground, the table tennis
ball is ejected upwards at high velocity. The IPT problem consisted in
two questions. \textsl{How high may a ping-pong ball jump using the
  setup on the video? What is the maximal fraction of the total
  kinetic energy that can be transferred to the ball?}~\cite{ipt2020}

French undergraduate students in the third year of university do not
have a background in hydrodynamics nor in capillarity. Most of the
first twenty hours of theoretical investigation were therefore devoted to
learn in autonomy the basic concepts in this field. With the help of
the supervising team, the project team produced conservation arguments
based on energy, as suggested by the IPT problem formulation, and on
momentum transfer. We summarise below their theoretical elaboration,
rephrasing them in a more rigorous scientific language. It must not be
understood as a final answer to the problem, but as an intermediate
stage aiming to prepare the experiment and to identify the relevant
control parameters.
 \begin{figure}[tbp]
\includegraphics{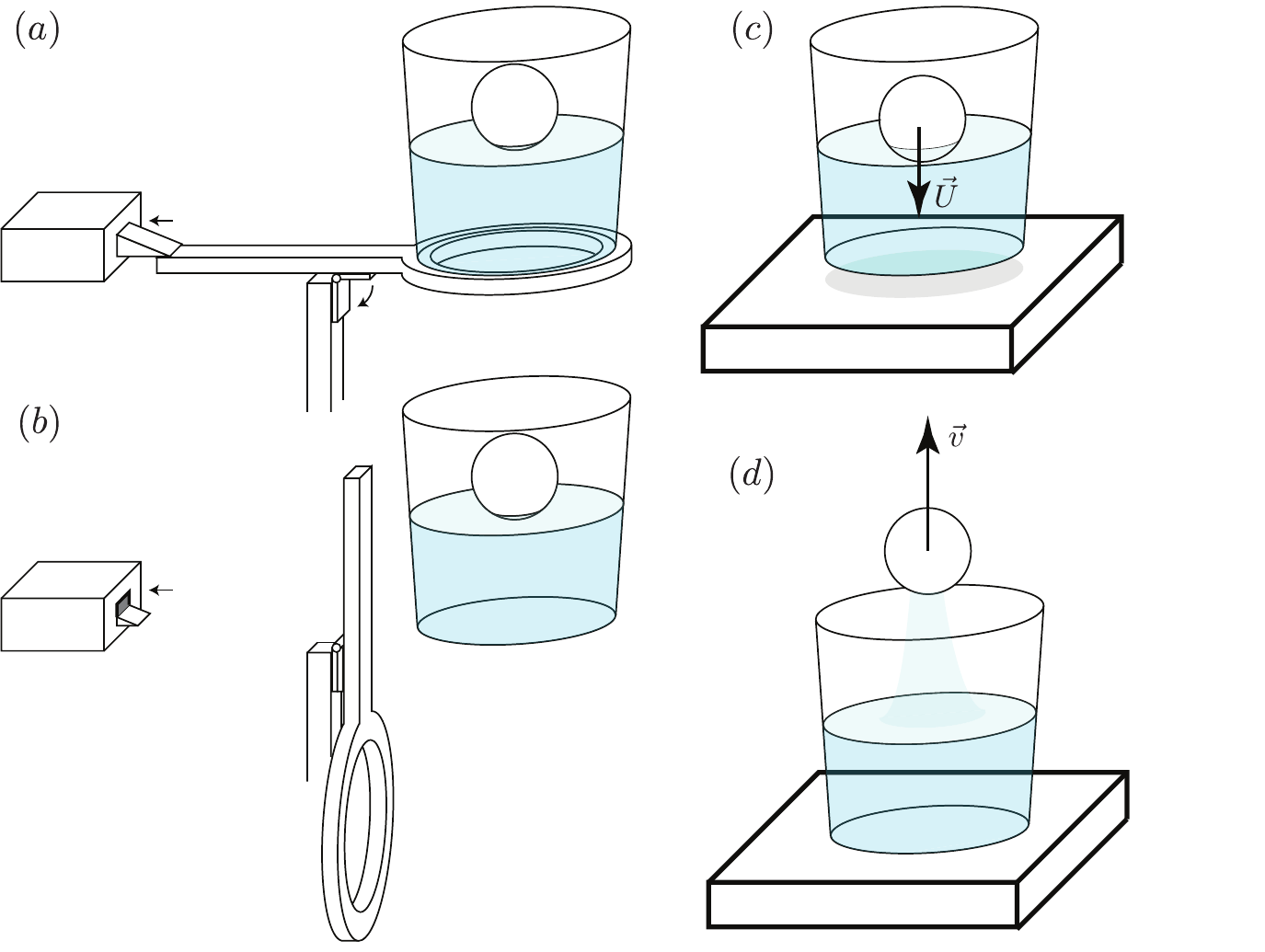}
\caption{Schematic of the experimental set-up. (a) The glass with water and table tennis ball rests on
  a trapdoor mounted on a spring-loaded hinge. (b) When the lock's
  bolt retracts, the trapdoor is released and pivots, moving away
  from under the glass at more than gravitational acceleration due to
  the spring. The glass starts its free fall. (c) Glass, water and
  ball impact the flat ground at a velocity $\vec U$. (d) Shortly after the
  impact, the ball is ejected upwards at velocity $\vec v$.}
 \label{fig:schematicCanon}
 \end{figure}

\subsection{The energetic upper bound}
\label{sec:energy}
A plastic cup of mass $m_c$ is filled by a mass $M-m_c$ of liquid. A
table tennis ball of mass $m \ll M$ is deposited at the surface, in the
centre of the cup. The cup is released from a height $\zeta$ above a flat
solid ground whose mass is much larger than $M$. It reaches the ground
with a velocity $U \simeq \sqrt{2g\zeta}$. After the impact, the table tennis ball
is ejected with a velocity $v$. The aim of the project, once rephrased
in scientific terms, is to determine the dependence of $v$ on the
control parameters of the experiment: the release height $\zeta$, the
amount of water in the cup, the size of the cup, the initial rotation
$\omega$ of the liquid, the surface tensions involving the liquid and the
ball.

The kinetic energy of the ejected table tennis ball $\frac 12 m v^2$ is
limited by the total initial gravitational energy $(M+m)g\zeta$, so the
ball's velocity has the following upper bound:
\begin{equation}\label{eq:upperbound}
v < \sqrt{\frac{M}{m}} U ,\qquad m\ll M
\end{equation}
Any change in wetted area of the ball during this process can be
neglected in the energy budget: the capillary energy difference
between a dry and a wet ball,
$4\pi R^2\gamma\cos\theta < 4\pi R^2\gamma \simeq \SI{3e-4}{\joule}$, where
$\gamma$ is the liquid-air surface tension and $\theta$ a typical contact angle
of the liquid on the ball (see Fig.~\ref{fig:SFigAlcohol}), is orders of
magnitude below the kinetic energy, of around \SI{2}{\joule}
(the inverse ratio is the Weber number
$\mathrm{We} = \rho_w U^2R/\gamma \simeq \num{1e4} \gg 1$). Finally, the large
Reynolds number $\mathrm{Re} = RU/\nu \simeq \num{1e5}$ implies that inertia dominates over
viscous dissipation.

Capillary forces alone cannot be responsible for the ball
ejection itself. A rough order of magnitude shows that the work they
can perform on the ball could accelerate it to at most
$\sqrt{\gamma R^2/m} \simeq \SI{0.2}{\metre\per\second}$. Although some plants
and mushrooms are known to use capillary forces to eject
spores\cite{ingold1939spore,noblin2009surface,dumais2012vegetable}, a
capillary `snap' mechanism due to sudden meniscus shape change during
impact can be ruled
out here. This reflects in the high Weber number estimated above.

\subsection{Scaling law based on momentum conservation during the
  shock}
\label{sec:momentum}
Imagine the situation where a glass partly filled with water, without
the table tennis ball, hits the ground flat. Assume the free surface
of the water is horizontal. The impact sees a pressure shock wave
propagating through the water from the bottom upward, progressively
stopping the water motion. At the shock front each water layer
effectively collides with the cylinder of water below which is already
at rest. By invariance in the horizontal plane, if menisci and free
surface perturbations can be neglected, neither wave nor central jet
is excited: water goes to rest, the free surface remaining
flat~\cite{antkowiak2007}.

The table tennis ball displaces water, but in terms of momentum it can
be replaced by an equal mass of water. If the table tennis ball were
at its equilibrium buoyancy position at the moment of impact, this
substitution would precisely take us back to the case of a flat water
surface (see next section), so we would not expect the ball to rebound
any more than the substituted water.\footnote{This neglects the balls
  elasticity which can store some potential energy during impact,
  causing the ball to rebound almost as on solid ground, at least for
  weak immersion depth from which the ball can extract with only
  moderate loss of momentum.} However pictures just before impact
reveal that the ball position differs from static equilibrium, in
which case the following argument shows that ejection occurs.

Consider a point inside the liquid at depth $h$. The pressure
increases from atmospheric pressure by about $\rho U c$ when it is
reached by the stopping front, whose unknown velocity\footnote{In a
  hard-walled glass on a hard floor the front is expected to be
  relatively localised, and its speed close to the bulk sound speed.
  For more compliant vessels the stopping front should be broader and
  the pressure increase slower.} is noted $c$. This excess pressure is
maintained until the stopping front has reached the free surface and a
relaxation front has propagated backwards, and its integral over time
is $\rho U h$, corresponding to the momentum surface density above the
considered point. Integrated across the immersed surface of the table
tennis ball and the shock duration, this pressure injects an upwards
momentum $m_wU$ into the ball, where $m_w$ is the mass of water
displaced by the ball, that may overcompensate the downward momentum
$mU$ of the ball before impact. The net momentum transfer is therefore
expected to eject the ball at a speed
\begin{equation}
v = \frac{m_w-m}{m}\;U 
\label{eq:predic}
\end{equation}
 \begin{figure}[tbp]
\includegraphics{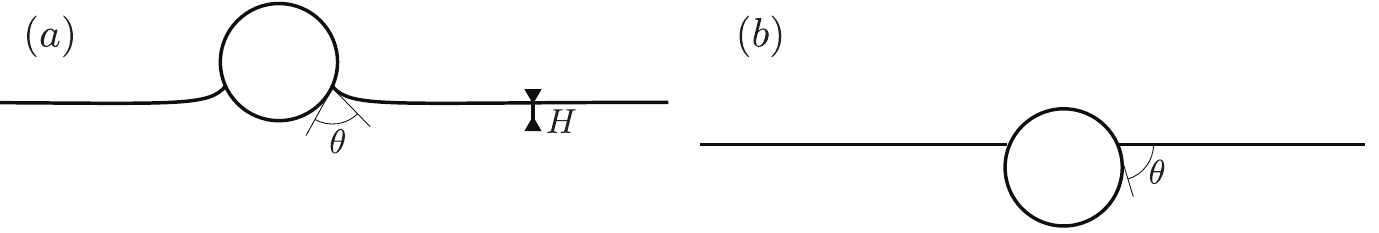}
 \caption{Schematic showing the equilibrium position of the ping-pong ball (a) under gravity  and (b) during the free flight.}
 \label{fig:SchematicEquilibrium}
 \end{figure}
 
\subsection{Equilibrium position of a table tennis ball with and
  without gravity}
\label{sec:immersion}
At first sight, the previous formula seems to imply that no ball
ejection is to be expected. Indeed, static balance between Archimedes
buoyancy force and gravity states that $m_w=m$ for a ball floating on
a static bath (Fig.~\ref{fig:SchematicEquilibrium}a). Neglecting
surface tension, the immersed volume under gravity is related to the
depth $H$ by:
\begin{equation}
\frac{m}{\rho} =  \pi H^{2} \left(R-\frac H3\right) 
\end{equation}
where $R$ is the ball radius and $\rho$ the fluid density.
%
However, here, the ball during the free flight does not feel gravity.
Its equilibrium position and therefore $m_w$ is controlled by surface
tensions. The interface must obey Young's law selecting the contact
angle both on the ball and on the plastic glass. With no hydrostatic
pressure gradient during free fall the interface connecting the two
circular contact lines on the centred ping-pong ball and on the glass
is a constant mean curvature surface, or Delaunay
surface~\cite{delaunay1841surface}. The Laplace pressure associated
with this constant curvature balances the capillary forces on the
glass and on the ping-pong ball. When the ball is in equilibrium, the
resultant of capillary forces and Laplace pressure acting on it must
vanish. A zero Laplace pressure configuration arises in the simplest
case of a contact angle equal to \SI{90}{\degree} on the glass. The
interface is simply flat and horizontal
(Fig.~\ref{fig:SchematicEquilibrium}b), and joins the ball
horizontally \cite{Finn06,Finn08,Marchand2011a}. Young's law selects
the ball immersion depth. Then, the volume displaced corresponds to
$H=R(1+\cos \theta)$ leading to:
\begin{equation}
m_w = \rho R^{3} \left(1-\frac{\cos \theta}3\right) \cos^2 \theta
\end{equation}
The calculation of the displaced mass $m_w$ from the general unduloid
or nodoid interface, as a function of the contact angle on the glass,
remains outside the scope of this paper.
\begin{figure}[tbp]
  \includegraphics{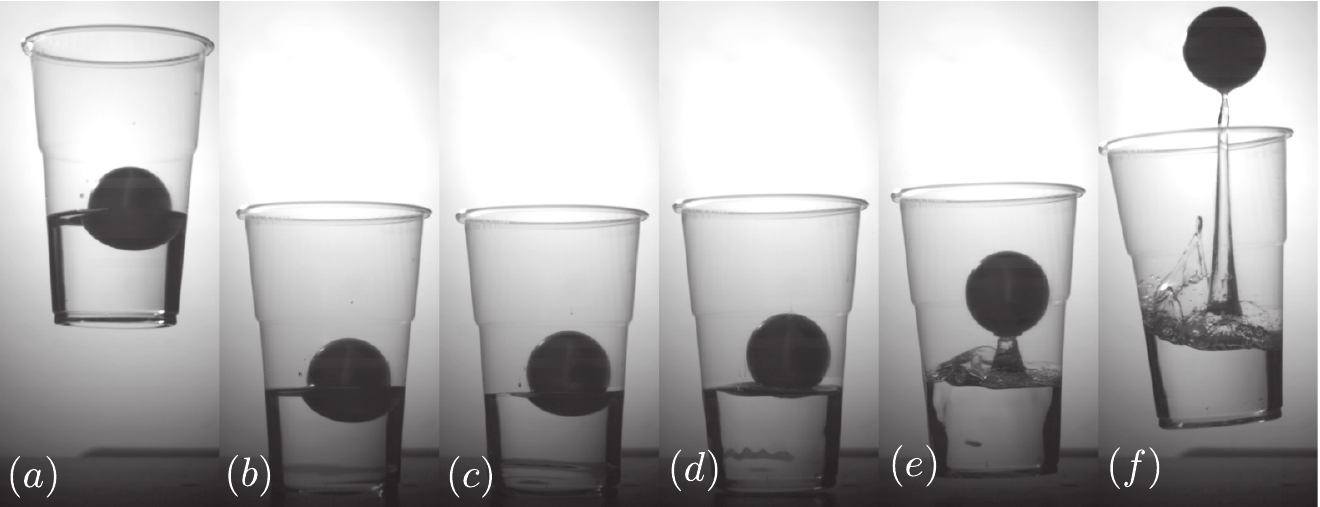}
  \caption{Series of pictures extracted from a fast movie. Images
    taken at $t=$ \SIlist{0;16;17;18;21;33}{\milli\second}. The
    diameter of the table tennis ball is \SI{40}{\milli\meter}.}
  \label{fig:SFigSerie}
\end{figure}
 
\section{Experimental results}
\subsection{Experimental set-up and preliminary observations}
The first goal of the project has been to perform controlled
experiments. In order to test the theoretical ideas, a first set-up
was designed to strike a static glass from below. A large metallic
block, used as the mass of a meter scale pendulum attached to a
spring, formed a `hammer' that would hit the plastic glass, containing
a table tennis ball floating on water, through the central hole of an
annular support (Fig.~\ref{fig:schematicHammer}). The geometry was
adjusted to get a planar collision between the hammer and the glass
bottom. Except for the case where the water was stirred to make it
  rotate, or for the case of a ball in an empty glass, without water,
the table tennis ball was never ejected from the glass. This is
exactly what is expected from equation \eqref{eq:predic} for
$m_w=m$ (for why rotation changes the outcome radically see
section~\ref{sec:surfactantrotation}). In conclusion, the free flight
is essential to the cannon effect, which implies that the conditions
under which the glass is released at null velocity must be controlled
with precision.

In the final set-up, a spring-loaded trapdoor supports the plastic
glass (Fig.~\ref{fig:schematicCanon}). An electric lock is used to
trigger the release of the trapdoor, which is hollowed out beneath the
glass to avoid aerodynamic suction and disturbances as much as
possible. The spring is chosen such that the trapdoor moves out of the
way faster than the glass falls. A visco-elastic shock absorber made
of a polymeric foam prevents a rebound of the trapdoor once it reaches
a vertical orientation. The trapdoor is carefully levelled, to within
$0.1^\circ$ of the horizontal, so that the glass is upright when released.
Within this precision, thanks to the elasticity of the plastic, the
impact takes place over the whole surface of the glass bottom.

The glass impact is imaged with a $1632\times1200$ fast video camera
borrowed from the laboratory \textsl{Matière et Systèmes Complexes}.
The resolution is \SI{190}{\micro\meter} per pixel in the focal plane
and the imaging frequency is \SI{1000}{\hertz}. The impact velocity
$U$ and the ball ejection velocity $v$ are automatically measured from
the slope of a space time diagram (strip-videograph) of a line passing
though the centre of the ball.

Figure~\ref{fig:SFigSerie} shows a sequence of images in a typical
experiment. The motion of the table tennis ball reverses direction
less than a millisecond after the impact. The ejected ball is always
accompanied by a columnar liquid jet. We checked that there is no such
central jet in the absence of the ball.

\subsection{Dependence of the ejection velocity on the release height}
\begin{figure}[t!]
\includegraphics{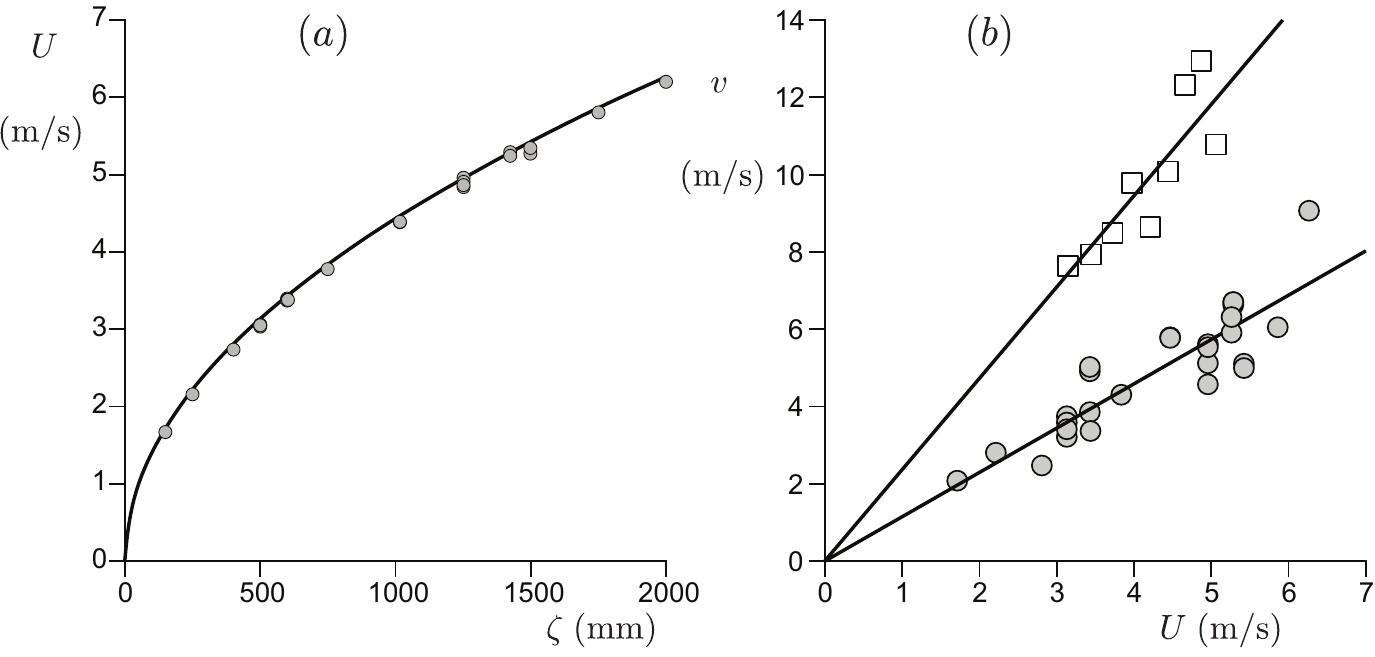}
\caption{(a) Dependence of the impact velocity $U$ on height $\zeta$. The
  solid line corresponds to $\sqrt{2g\zeta}$. (b) Ejection velocity
  $v$ as a function of impact velocity $U$, varying the release height
  $\zeta$. Circles: without rotation. Squares: with a constant rotation.}
 \label{fig:VelocityHeight}
\end{figure}
We first consider the dependence of ejection velocity $v$ as a
function of the impact velocity $U$. Figure~\ref{fig:VelocityHeight}(a)
compares the relation between $U$ and the release height $\zeta$ to the
conservative theory, neglecting the drag force exerted by air. Its
overall agreement is excellent and one can judge the quality of the
velocity measurement.

We performed two series of experiments using the same type of plastic
glass and the very same table tennis ball. In the first series the fluid
does not rotate, while in the second the water is stirred manually
beforehand. We tried to reproduce the same initial condition when
depositing the table tennis ball at the surface of the water (wetting
meniscus height, centring in the glass). In many cases, the glass
broke at the impact, in which case the velocity is much smaller than
that reported in figure~\ref{fig:VelocityHeight}(b). Those cases were
discarded. The dispersion of the data presented in this figure mostly
result from the contact angle hysteresis and on the exact impact
conditions. Despite dispersion, one can conclude that the ejection
velocity is reasonably linear in impact velocity, both with and
without rotation.

\subsection{Dependence of the ejection velocity on the ball mass}
 \begin{figure}[t!]
\includegraphics{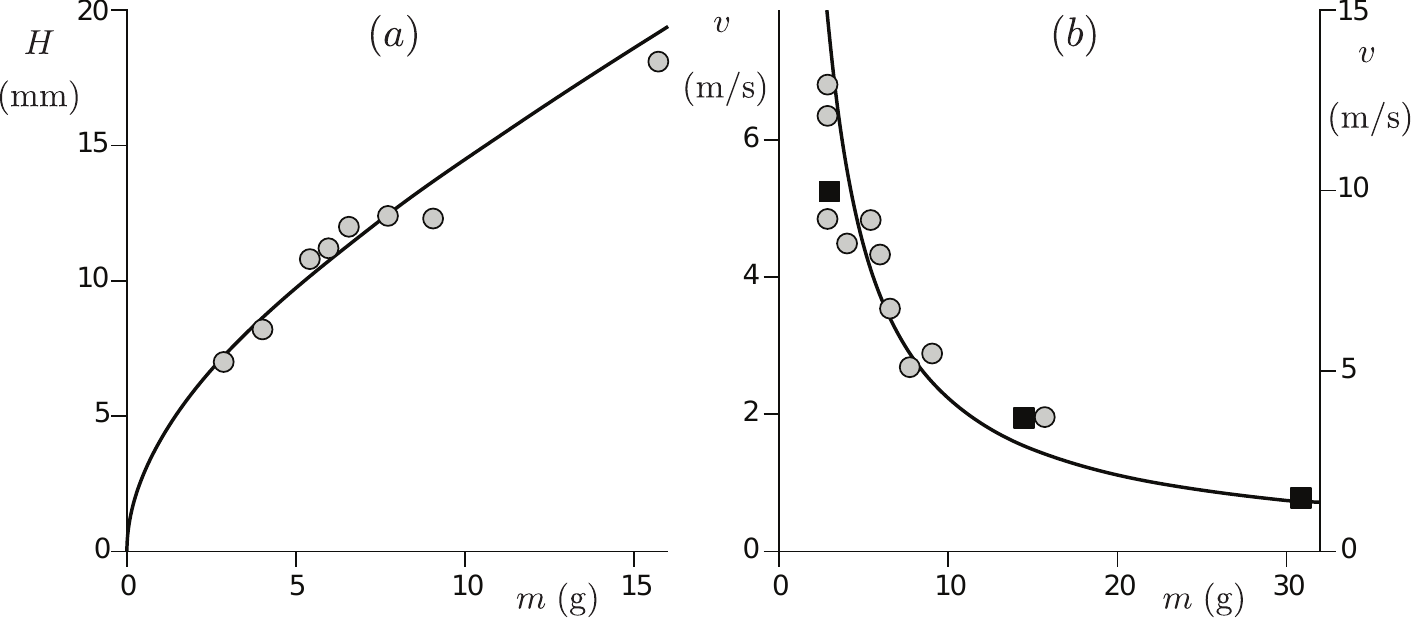}
\caption{(a) Depth $H$ as a function of the ball mass $m$ in the
  static case. The solid line is the theoretical prediction, when
  capillarity is neglected: $m=\pi \rho_w H^2 (R-H/3)$. (b) Ejection velocity
  $v$ as a function of the ball mass $m$, with (black squares, right
  axis, release height $\zeta= \SI{1}{\metre}$) and
  without (circles, left axis, release height $\zeta= \SI{1.42}{\metre}$) rotation.}
 \label{fig:SfigWeigth}
 \end{figure}
 We have changed the ball mass by injecting some liquid inside through
 a small hole by means of a hypodermic needle.
 Figure~\ref{fig:SfigWeigth}(a) shows the immersion depth of the ball
 as a function of the ball mass $m$, as compared to the theory
 neglecting surface tension. The agreement is, as expected, very good.
 Figure~\ref{fig:SfigWeigth}(b) shows the ratio of the ejection
 velocity to the impact velocity as a function of the ball mass $m$.
 The data obey a very simple relationship: the momentum transferred to
 the ball, $m v$ is roughly constant. Note that this is not entirely
 consistent with the prediction of equation \eqref{eq:predic}.

\subsection{Dependence of the ejection velocity on surface tension and rotation}\label{sec:surfactantrotation}
\begin{figure}[tbp]
\includegraphics{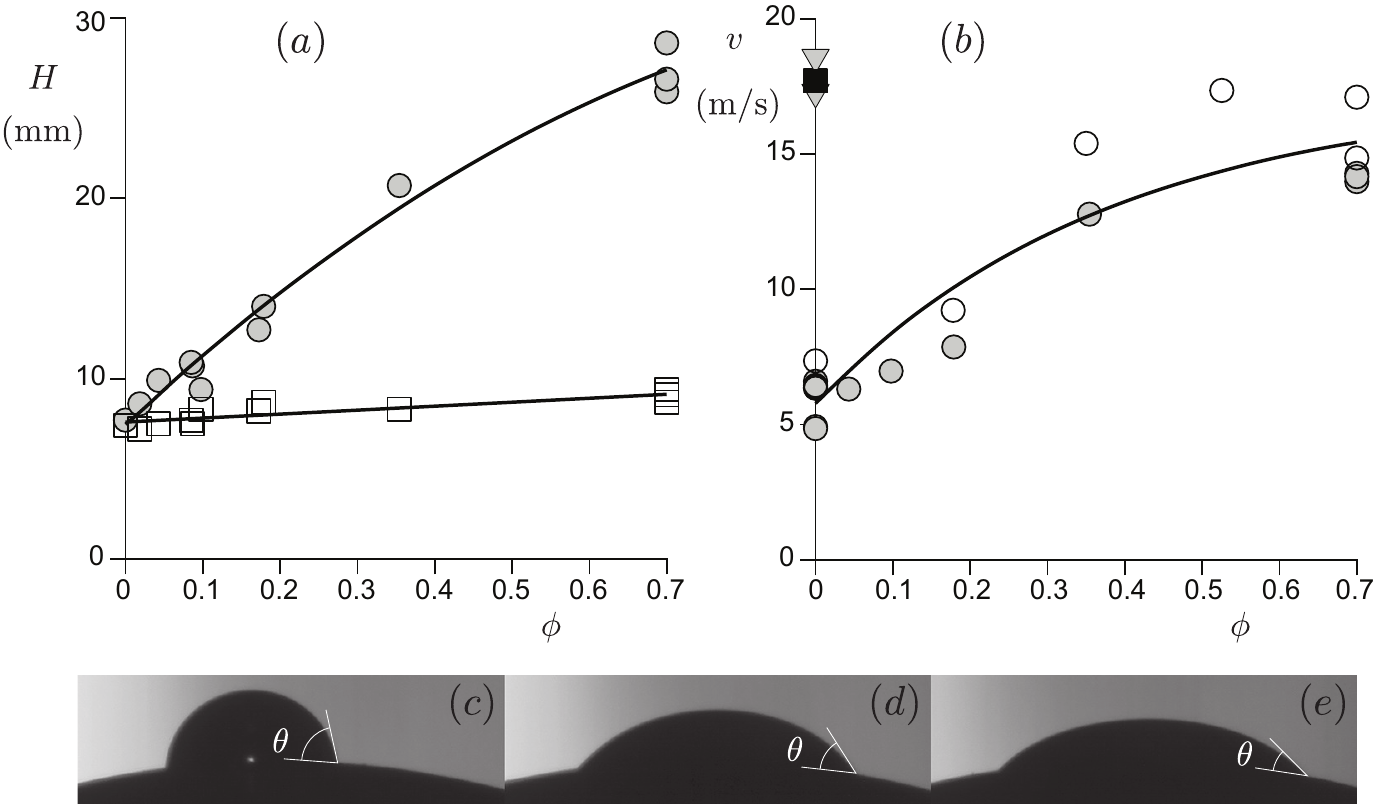}
\caption{Dependence on surface tension determined using a mixture of
  water and ethanol. (a) table tennis ball depth $H$ as a function of
  the ethanol fraction $\phi$. Circles: just before impact (free fall
  height $\zeta=\SI{1.42}{\metre}$); squares: at rest. (b) Circles:
  Ejection velocity as a function of the ethanol fraction $\phi$, release
  height $\zeta=\SI{1.42}{\metre}$. The grey and white symbols correspond
  to two types of plastic glass. Square: water with surfactant,
  $\zeta=\SI{1.06}{\metre}$. Triangles: water with a strong rotation,
  $\zeta=\SI{1.41}{\metre}$. The solid lines in (a) and (b) are
  phenomenological parabolic fits. (c) Drop of liquid on the table
  tennis ball: from left to right, water, $\phi=\SI{35}{\percent}$
  ethanol and $\phi=\SI{70}{\percent}$ ethanol.}
 \label{fig:SFigAlcohol}
 \end{figure}

Two independent factors can greatly increase the ejection velocity.
Stirring the water into rotation --- a trick used in the introductory
video to centre the table tennis ball --- has the effect of creating a
radial pressure gradient that will drag the ball into the liquid
during free fall. Likewise, increasing the wettability of the ball's
surface will cause surface tension to pull the ball under during free
fall. In both cases the net result is an increase in displaced volume,
and a strongly increased ratio of ejection velocity to impact
velocity. To change wettability we have found two routes to be equally
effective, either adding surfactant to the water, or using a mixture
of water and ethanol to vary the liquid surface tension and therefore
the liquid contact angle $\theta$ continuously (Fig.~\ref{fig:SFigAlcohol}(c)).

Ethanol wets the table tennis ball plastic more than water. As a
consequence, the contact angle $\theta$ is lower. We have measured the
immersion depth of the ball in the static case, under gravity
(Fig.~\ref{fig:SFigAlcohol}(a)). When going from pure water to pure
ethanol, $H$ remains almost constant. Indeed, gravity is the dominant
force determining the equilibrium position of the ball, and the change
of the liquid density is small. On the opposite, the immersion depth
just before impact strongly increases with the volume fraction $\phi$ of
ethanol. This directly shows that the equilibrium of the ball changes
during the free flight: capillarity, which is negligible for the
equilibrium position under gravity becomes dominant in the
micro-gravity conditions of free fall. The lower the contact angle,
the deeper the ball gets immersed, leading to a stronger momentum
transfer during the collision.

Fig.~\ref{fig:SFigAlcohol}(b) shows the ejection velocity $v$ as a
function of $\phi$ for the same impact velocity
$U\simeq \SI{5}{\metre\per\second}$. With \SI{70}{\percent} ethanol, the
ejection velocity is three times higher than with water. Two series of
measurements performed on two different dates, with distinct types of
plastic glasses, present a systematic difference. This may either be
due to different wetting properties of the ball used, or point to an
influence of either geometric or material properties of the plastic
glasses. Plastic dissipation at impact may change the deceleration
kinetics. We found for instance that test experiments using paper
cups, whose hollow bottom deforms permanently at impact, produced much
lower restitution coefficients. The black square corresponds to a run
performed with a surfactant --- commercial dish-washing liquid --- which
causes water to wet the plastic ball. The ejection velocity is
observed to be higher than with ethanol, as expected, and reaches
\SI{50}{\percent} of the upper bound from eq.~\eqref{eq:upperbound}.
We have also added two runs performed in water with a strong rotation,
which also exhibit high ejection velocities (angular frequency of the
ball during free fall $\omega \simeq \SI{1.4}{\radian\per\second}$).

It may come as a surprise that the ball depth can change significantly
during the free fall phase, which lasts only about \SI{0.5}{\second}.
Indeed, the suction force on the ball is weak: in the rotation case
the pressure below the ball is roughly \SI{1}{\pascal} below
atmospheric pressure, causing a downward force of merely
\SI{1}{\milli\newton}. This force is small, but as the otherwise
dominant buoyancy forces are suspended during free fall, it can
accelerate the ball downwards and move it up to \SI{3}{\centi\meter}
neglecting added mass effects. This is in accordance with
observations, which show the ball almost completely submerged for
strong rotation rates. Capillary forces acting on the perimeter of the
ball are in the same milli-Newton range.

%

\section{Conclusion}
\subsection{A physical picture of the phenomenon}

The most striking observation on the video recordings is the change in
immersion depth of the ball compared to the static buoyancy
equilibrium position, which we attributed to the effect of either
capillary forces or vortex depression (in the case of rotating liquid)
acting during free fall. During impact the pressure inside the liquid
in excess of atmospheric pressure suddenly increases from zero to
large values (for rigid glasses on hard floors up to $\rho\,Uc$ where
$c$ is the speed of sound in the liquid). As a result the ball is
pushed up very strongly, and accelerates upwards to velocities in
excess of the impact velocity.

The physical picture of the dynamics during impact is not unique and
depends on the relative magnitude of several time scales that are
involved. One is the time it takes for the impact shock to travel
through the liquid, from the bottom to the free surface. In a rigid
glass this would be the height of the liquid free surface above the
glass bottom divided by the bulk speed of sound in the liquid: around
\SI{50}{\micro\second}. In a compliant vessel the elasticity of the
walls reduce the effective speed of the shock
front\cite{rubinow1971wave}. A second time scale is the impact
duration. On a hard substrate this is equal to the previous time, but
on a soft substrate (picture a rigid glass falling on a soft foam
layer) it may be governed by the total mass of the glass and the
elasticity of the substrate. Last, but not least, the impact duration
has to be compared to a third time scale, the time for the table
tennis ball to accelerate out of the liquid. For short impacts in this
scale the momentum transfer will be maximal, while for very long
impacts the ball leaves the liquid early with only part of the
momentum transfered.

It turns out that except for the very last case mentioned above, the
prediction of eq.~\eqref{eq:predic} is the same even outside the hard
glass hard bottom limit described when deriving it in
section~\ref{sec:momentum}. It is instructive to describe the momentum
transfer mechanism in other regimes. Consider the case where the
impact duration is much longer than the time required for a bulk
compression wave to propagate through the liquid from the bottom to
the free surface: as all water molecules then share the same velocity
and decelerate at the same rate, the pressure becomes a linear
function of depth just as in the hydrostatic case. Its gradient is
much larger however, because the deceleration rate is larger than
gravitational acceleration $g$ by a factor given by the ratio of free
fall duration (\SI{500}{\milli\second}) to impact duration
($\approx\SI{1}{\milli\second}$). During the brief impact period the
glass-water-ball system can thus be considered as subject to a high
effective gravity field, and it is clear that the ball will tend to
relax very quickly to its buoyancy equilibrium position. Arriving at
this position, having acquired momentum, it continues upward.

A precise calculation of the ejection velocity and the energy transfer
coefficient is undoubtedly a very complex hydrodynamic problem. The
students were not expected to arrive at a quantitative restitution
model. Their experiments do however confirm the expected
proportionality between $v$ and $U$, and provide the correct order of
magnitude. It also predicts the dependence on wetting. The prediction
of the restitution coefficient $(m_w-m)/m$ could not be tested
quantitatively, due to the difficulty to determine $m_w$
experimentally. This remains one of the major issues to tackle in the
future.

To answer the International Physicists' Tournament question on the
maximum kinetic energy transfer, one would have to better understand
the simultaneous creation of a liquid jet which carries part of the
momentum.

\subsection{Worthington jet vs dynamic entrainment by the ball}
Imagine the ball would magically disappear at the very instant the
glass impacts the floor. The void left by the ball would collapse and
create a jet exactly in the same way a meniscus on the side-walls of a
tube collapses and creates a jet upon
deceleration~\cite{antkowiak2007,lavrentiev1980}. This collapse
dynamics was also studied under standard gravitational acceleration in
cavities created by the impact of a solid body~\cite{gekle2010} or in
cavities left at the rupture of surface bubbles~\cite{boulton1993},
and the explanation of the basic jet formation mechanism dates back to
Worthington and Cole\cite{worthington1897,worthington1900}. The radial
inward motion induced by the collapse causes a strong dynamical
pressure to build up at the centre, by which the liquid is expelled
along the axis. The presence of the table tennis ball changes the
boundary condition at the cavity's free surface by adding a dynamic
pressure (the ball is injecting its momentum as it decelerates as
well). In fact, taking the table tennis ball out and filling the
cavity partly by this ball's mass in liquid should lead to the same
momentum balance. With a ball at its equilibrium buoyancy depth
($m_w=m$) this amounts to completely filling the cavity, and producing
no jet (an elastic ball still allows for rebound and jet). However
when the ball moves into the liquid we have an effective cavity and
the jet velocity is expected to scale as $(m_w-m)U/m$ just like the
ball velocity according to eq.~\eqref{eq:predic}. In practice the two
velocities appear to be approximately equal. If the jet were pushing
the ball we would expect to see a splash on the bottom of the ball,
which is rarely seen (and when it is, the mass involved is minute).

On the opposite, we could consider that the liquid jet is entrained by
the ball, slowing it down. When a solid object is withdrawn from a
bath of liquid in partial wetting, the contact line can remain steady
in the frame of reference of the lab if the solid velocity is
sufficiently small \cite{SADF07}. Above a critical value of the capillary number
$\eta v/\gamma$, a dynamical wetting transition occurs
\cite{BR79,Q91,SP91,E04b,SDAF06}: as capillary forces can no longer
compete with the viscous stress that develops inside the flow, liquid
is entrained dynamically \cite{RevSnoB13}. The velocity difference
between the contact line and the solid surface is selected by the
critical capillary number, and is proportional to $\gamma/\eta$. This may
explain the conical shape of the water column below the ball, similar
to that observed in the cat lapping problem~\cite{reis2010cats}.

If this interpretation is correct, the resistive force exerted by the
liquid on the ball would remain relatively small, scaling on
$\gamma R$. Indeed, at the scale of the ball, everything goes as if the effective
macroscopic contact angle was vanishing. The overall viscous force
exerted on the ball is then simply governed by the unbalanced Young
force. This does not preclude the existence of an added mass to the
moving sphere during the impact itself, scaling on $\rho_w R^3$.

Both the Worthington jet and the dynamic entrainment explanations
shall contain a part of the correct picture: on the one hand, the
pressure gradient inside the fluid should accelerate both the ball and
the liquid below; on the other hand, the flow structure close to the
contact line should resemble that observed at the critical point for
dynamical entrainment.

\subsection{``Phy Ex'' and research in macroscopic physics}
In conclusion, we have exemplified here one particular physics project
performed in the \textsl{Phy Ex} initially created by Yves Couder.
Indicative of the fruitful proximity of this teaching to
full-fledged research is that occasionally, a successful experiment
leads to a publication of the group of students and their supervisors
in a scientific journal\cite{daerr2019}. More famously some research
projects started in \textsl{Phy Ex} have lead to discoveries that
spawned a whole research branch over the following years. One example
is a problem involving Quincke rollers, proposed by Denis Bartolo when
he was teaching \textsl{Phy Ex} as faculty member of Université Paris
7, which became the starting point of his subsequent work on the
collective behaviour of large collections of these Quincke
rollers\cite{bricard2013emergence,bricard2015emergent}. A second
example is Yves Couder's proposal of a problem on the Faraday
instability, cited above, that led to the fortuitous discovery of
walking droplets, which have since been largely studied in
laboratories around the
world\cite{couder2005walking,couder2005bouncing,bush2010quantum,bush2018introduction}.

\bibliographystyle{crunsrt}


\bibliography{CrasPingPong}

\end{document}


\centerline{\textbf{\textsf{\Large Supplementary movies}}}

\begin{description}

\item[SM\_movie1] 
  Ping pong ball water cannon in reference conditions (see main text
  fig.~4 for snapshots): Cup mass $m_c=\SI{6.2}{\gram}$, cup+water
  mass $M=\SI{150}{\gram}$, table tennis ball mass
  $m=\SI{2.84}{\gram}$, drop height $\zeta=\SI{1.411}{\metre}$. Scale bar
  and time code are overlayed on the movie.

\item[SM\_movie2] 
  Liquid is \SI{70}{\percent} ethanol, other conditions as in
  SM\_movie1, in particular $M=\SI{150}{\gram}$. The modified wetting
  conditions on the ball cause its enhanced immersion at the end of
  the free fall, and a more than threefold ejection velocity.

\item[SM\_movie3] 
  Conditions as in SM\_movie1, but the water was stirred into rotation
  before releasing the glass. Note again the more immersed position of
  the ball just before impact and higher ejection speed, compared to
  SM\_movie1.

\item[SM\_movie4] 
  Cup mass $m_c=\SI{15.7}{\gram}$, cup+water mass $M=\SI{150}{\gram}$,
  table tennis ball mass $m=\SI{2.81}{\gram}$, drop height
  $\zeta=\SI{1.057}{\metre}$. Here the glass falls onto a
  \SI{10}{\centi\metre} thick rubber foam, increasing the duration of
  its rebound to about \SI{20}{\milli\second}. Note how the ball is
  ejected in a fraction of that time. When the glass' velocity becomes
  zero (corresponding to an overlay time stamp of
  $t=\SI{0}{\milli\second}$), the ball has already left the liquid
  entirely. It is subsequently overtaken and engulfed by the
  Worthington jet, which is accelerated over the whole impact
  duration. This contrasts with the other movies where ball and jet
  tip velocities appear to be equal.

\end{description}